\documentclass{article}
\usepackage[utf8]{inputenc}
\usepackage{amsmath}
\usepackage{amssymb}
\usepackage{listings}
\usepackage{comment}
\usepackage{xcolor}
\usepackage{graphicx}
\usepackage[round]{natbib}
\usepackage{hyperref}

\title{Efficient implementation of sets and multisets in R using hash tables}
\author{Giacomo Ceoldo, Ernst C. Wit}
\date{}

\begin{document}

\maketitle

\begin{abstract}
The package \texttt{hset} for the R language contains an implementation of a S4 class for sets and multisets of numbers.
The implementation, based on the hash table data structure from the package \texttt{hash} \citep{hash}, allows for quick operations when the set is a dynamic object.
An important example is when a set or a multiset is part of the state of a Markov chain in which in each iteration various elements are moved in and out of the set.
\end{abstract}

\section{Introduction}
Sets are the most basic and fundamental containers of objects in mathematics.
According to set theory (almost) all objects in mathematics are, or can be described as, sets. 
Some objects have additional mathematical and computational structure, such as multisets, lists, vectors, stacks, etc.
Sets and multisets have been less developed in programming languages than other objects, such as vectors and lists.
The reason is that the latter ones are used very frequently in algorithms, so almost all programming languages have a built-in implementation of them.
These implementations do not reflect the mathematical derivation of these objects from set theory, as their structure allows the use of much more computationally efficient implementations and algorithms.
Sets and multisets as basic containers are nevertheless very important, especially for discrete probabilistic and statistical models.
The aim of the paper is to provide an efficient implementation for algorithms that use sets and multisets as containers.
Our implementation is based on the \texttt{hash} package \citep{hash}.
It is efficient because the hash table data structure allows search, insertion and deletion of one element in the table in constant time.

The mathematical definition of sets and multisets is given in Section \ref{sec:setsAndMultisets}, where our implementation and its semantic is also discussed.
In Section \ref{sec:sets algebra}, relations and operations between sets and multisets are defined mathematically, and their implementation is described.
The performance of our implementation is discussed in Section \ref{sec:performance}. 
In Section \ref{sec:application} an application of our implementation of sets and multisets as states of a Markov chain is provided.

\section{Sets and multisets}
\label{sec:setsAndMultisets}

Sets, multisets and some other constructions/containers derived from them, are introduced mathematically in Section~\ref{subsec: mathematical definition - sets and multisets}.
The emphasis is on how containers differ on how much structure is ``imposed" on them.
In sets elements are either in or out of them, in multisets it is also relevant how frequently an element is contained, for sequences the order of the elements is also important. 
The R package \texttt{hset} is introduced in Section~\ref{subsec: computational implementation hset}, where it is also discussed which objects can be included as elements, and how they are stored in a data structure based on the package \texttt{hash}.
The semantic of \texttt{hset} objects is described in Section~\ref{subsec: semantic of hset - sets and multisets}, with some other functions that are used to control these objects, that depend on the chosen semantic.

\subsection{Mathematical definition}
\label{subsec: mathematical definition - sets and multisets}
\emph{Sets} are defined as collections of objects called \emph{elements}, or \emph{members}.
\emph{Set theory} can be used as foundation of mathematics. 
The existence of the \emph{empty set} $\emptyset=\{\}$ is postulated, $\emptyset$ is the only element such that 
\begin{equation}
\{a,\emptyset\} = \{a\}, \ \forall a,
\end{equation}
where equality between sets will be formally defined in Section \ref{subsec:relations - sets algebra}.
All elements of a set are considered to be sets themselves, and (almost) all objects in mathematics are constructed as sets.

An important example is the set $\mathbb{N}$ of \emph{natural numbers} that can be defined recursively as
\begin{equation}
0=\emptyset, \ 1=\{0\}=\{\emptyset\}, \ 2=\{0,1\}=\{\emptyset,\{\emptyset\}\}, \ ...   
\end{equation}
Another important example is the \emph{Cartesian product} of two sets, that is the set of ordered pairs with first element from $X_1$ and second element from $X_2$, defined as 
\begin{equation}
    X_1\times X_2 = \{X_1,\{X_2\}\},
\end{equation}
which contains elements $(a_1,a_2)$, for $\emptyset \ne a_1\in X_1$ and $\emptyset \ne a_2 \in X_2$.
Note that $\emptyset \times X = X \times \emptyset = \{\emptyset\}$ for all $X$, and that $(a_1,a_2)\ne (a_2,a_1)$ for $a_1\ne a_2$ (the Cartesian product is not \emph{commutative}).

Order and multiplicity of the elements of a set is not defined, that is
\begin{equation}
\{a,b\} = \{b,a\} = \{a,a,b\}, \ \forall a\ne b.
\end{equation}
\emph{Multisets} are defined as collections of elements with \emph{multiplicities}, that are non-negative numbers. 
We will always assume that the multiplicities of the elements are finite.
The order of the elements is not defined, but multisets of elements with different multiplicities are different:
\begin{equation}
\{a[m],b[n]\}\ne\{a[n],b[m]\}, \ \forall n\ne m, \ \forall \emptyset\ne a\ne b \ne \emptyset 
\end{equation}
If the multiplicity of an element is $0$, then the element is not contained in the multiset:
\begin{equation}
\{a[m],b[0]\} = \{a[m]\}, \ \forall a,b.
\label{multiset.with.element.with.zero.multiplicity}
\end{equation}

A set can be considered equivalent to a multiset of the same elements, all with multiplicity $1$:
\begin{equation}
X=\{a,b\} \cong \{a[1],b[1]\}=Y(X), \ \forall a,b,
\label{conversion.set.to.multiset.Y(X)}
\end{equation}
The injective function $X\mapsto Y(X)$ converts a set to the ``equivalent" multiset, but when applied to a multiset, it is the identity function: $Y=Y(Y)$.
The surjective function $Y\mapsto X(Y)$ maps the multiset $Y$ into its \emph{support}, that is the set $X(Y)$ of its elements, if $X$ is a set $X=X(X)$.
The size of a multiset (or of a set), is the number of its elements, that is 
\begin{equation}
\mathrm{size}(Y)=|Y|=|X(Y)|\in\mathbb{N}.
\label{size}
\end{equation}
The cardinality of a multiset $Y=\{a_i[m_i]\}_i$ is the sum of its multiplicities, that is
\begin{equation}
\mathrm{card}(Y)=||Y||={\textstyle\sum}_i m_i\in[0,\infty),
\label{cardinality}
\end{equation}
while for a set $X$, $\mathrm{size}(X)=\mathrm{card}(X)$.


Other constructions that will be used later on are based on the Cartesian product, that is used to define \emph{powers} of the set $X$ as
\begin{equation}
    X^k = X \times X \times ... \times X,
\end{equation}
for $k\in\mathbb{N}$, where $X$ is repeated $k$ times in the right hand side, $X^0=\emptyset$ and $X^1=X$.
Powers of $X$ are used themselves to define finite dimensional \emph{sequences} (or \emph{strings}) of elements in $X$ as
\begin{equation}
    A = (a_1, ..., a_l) \in X^* = \biguplus_{k\ge 0} X^k,
    \label{eq:sequences}
\end{equation}
so a sequence of length $l$ with elements in $X$ is an element of $X^l$, and the set of sequences $X^*$, that is the \emph{disjoint union} of all powers, contains all sequences of all finite possible lengths, including the \emph{empty sequence} $()$.
Sequences have even more structure than multisets, as they are ordered, meaning that $(a_1,a_2,...)\ne (a_2,a_1,...)$ for $a_1\ne a_2$.

Finite dimensional sequences are introduced because they are one of the most used objects in programming, so most languages have efficient built-in implementations of sequences with finite length.
However these implementations are often inefficient when a sequence is modified locally by operations of inclusion or removal of elements from the list.

In R finite dimensional sequences are implemented as objects of type \texttt{vector}, with sub-types \texttt{atomic} and \texttt{list} \citep[Chapter 3]{wickham2019advanced}.
In the next sections, our R implementation of sets and multisets is discussed in detail.

\subsection{Computational implementation}
\label{subsec: computational implementation hset}
In mathematics, elements of sets are usually considered sets themselves. 
Then a formal definition of elements is redundant, once sets are defined.
On the other hand, when sets are viewed as computational objects, a definition of elements is required because the elements are objects with, possibly various, data types stored in memory.

Sets and multisets are implemented in the \emph{R package} \texttt{hset}, as objects of \emph{S4 class} \texttt{"hset"}.
These objects are containers of elements, that are either numbers, or sets of numbers.
A more formal definition of objects that are valid elements will soon be given.
An object of class \texttt{"hset"} contains two \emph{slots}.
The main one, called \texttt{@htable}, is an hash table from the package \texttt{hash} \cite{hash}.
The second slot, called \texttt{@info}, is of class \texttt{"environment"}, which contains a Boolean value that distinguish sets from multisets.
The reason why the second slot is a ``trivial" (with one object) environment, rather than an object of class \texttt{"logical"}, is because environments and (logical) vectors have a different semantic, then it would be difficult to reason about ``sameness" of objects (sets or multisets).
The constructor for objects of class \texttt{"hset"} will be described at the end of this section, the semantic of our implementation will be discussed in detail in the next section.

The set $\texttt{S}$ of possible sets that can be stored is recursively defined as
\begin{equation}
\begin{split}
&X = \{a_1,a_2,...\} \in \texttt{S},\\
&a_i \in \texttt{S} \uplus \texttt{N},
\end{split}    
\end{equation}
so the element $a_i$ can be either a set, or a value in $\texttt{N}$ that is the set of \texttt{numeric} vectors of length $1$, without the values \texttt{Inf}, \texttt{-Inf}, \texttt{NaN}, \texttt{NA}, \texttt{NA\_integer\_} and \texttt{NA\_real\_}, that are excluded.
The symbol $\uplus$ in $\texttt{S} \uplus \texttt{N}$ is used to denote the disjoint union between the sets $\texttt{S}$ and $\texttt{N}$. 
The inclusion relation $\in$ between an element, and a set or multiset, will be formally defined in Section \ref{subsec:relations - sets algebra}.

The set $\texttt{M}$ of possible multisets that can be stored is recursively defined as
\begin{equation}
\begin{split}
&Y = \{a_1[m_1],a_2[m_2],...\} \in \texttt{M},\\
&a_i \in \texttt{S} \uplus \texttt{N},\\
&m_i \in \texttt{N}_{\texttt{+}},
\end{split}    
\end{equation}
where $\texttt{S}$ and $\texttt{N}$ are defined as above, and $\texttt{N}_{\texttt{+}}\subset \texttt{N}$ is the subset of values of $\texttt{N}$ that are strictly positive.
Note that in our current implementation multisets can not be elements of a set or a multiset, and that recursion occurs in the definition of $\texttt{M}$ because it occurs in the definition of $\texttt{S}$.
The \texttt{numeric} datatype includes \texttt{integer} and \texttt{double} as subtypes, so the elements can be also of these two types.
Vectors of type \texttt{numeric} of length 0, together with the \texttt{NULL} object, are considered equivalent to the empty set, so they are not included in an object of class \texttt{"hset"} (even though formally an empty set is included in every set).
Instead, \texttt{numeric} values of length at least $2$ and \texttt{list} values of every length are converted to elements of $\texttt{S}$, that is to sets, before being included as elements.

The package \texttt{sets} \citep{sets} uses a different approach, the sets in this library, that can be of classes \texttt{set} (sets), \texttt{gset} (\emph{generalized} sets), \texttt{cset} (\emph{customizable} sets), can contain elements of every type.
Two elements can be considered the same only if they have the same class, so for example the sets $\{ \texttt{2},\texttt{2L}\}$ and $\{\texttt{2}\}$, are not equal (the former has two elements), as \texttt{2} and \texttt{2L} have class \texttt{"numeric"} and \texttt{"integer"} respectively.
As a result, although our implementation is more limited because we only take into account sets and multisets of numbers, it is still somewhat closer to the mathematical definition in which $\texttt{2L}$ and $\texttt{2}$ represent the same number.

The \emph{hash table} implemented in \texttt{hash} is a data structure that contains \emph{key-value pairs}.
The \emph{keys} are different objects of type \texttt{character}, they are unique labels of pointers to the \emph{values}, that can be objects of every type except \texttt{NULL}.
The advantage of using an hash table to implement sets and multisets is that various operations that are: adding and removing a key-value pair from the table, checking to see if a key is present in the table, and returning the value associated with a key; only require on average a constant number of elementary operations.

There is an injection $k:\texttt{S} \uplus \texttt{N}\to \texttt{C}$, where $\texttt{C}$ is the set of character vectors of length 1, that is used as set of keys to label uniquely each possible element of a set or a multiset.
For example, the element
\begin{equation}
a_i=\{\texttt{-1},\texttt{1},\texttt{1L},\texttt{\{\}},\texttt{2},\texttt{11},\texttt{\{2,\{3\}\}}\},
\end{equation} 
is mapped to
\begin{equation}
k_i = k(a_i)=\texttt{"\{-1,\{\},\{2,\{3\}\},1,11,2\}"},
\end{equation}
where the "sub-elements" $\texttt{1}$ and $\texttt{1L}$ are both mapped to $\texttt{"1"}$, and the components of the character $k_i$ are in \emph{lexicographic order}, which has the mathematical properties of a \emph{total order}.
Note that the order is important to guarantee that $k$ is an injection.
For sets, all values of the hash table are equal to the empty character $\texttt{""}$, that is $v(k_i) = \texttt{""}$ for all $i$, whereas for multisets $v(k_i) \in \texttt{N}_{\texttt{+}}$.

Sets and multisets are created with the \emph{constructor} \texttt{hset} with three arguments that are \texttt{members}, \texttt{multiplicities} and \texttt{generalized}.
The first two arguments are \texttt{NULL} by default (empty set), the last one is \texttt{FALSE} by default.
If the second input is not \texttt{NULL}, \texttt{generalized} is set to \texttt{TRUE} if it was not the case.
Then it is checked that \texttt{members} and \texttt{multiplicities} are of the correct type, that they are coherent with themselves, and then they are included in the hash table.
The function \texttt{is.hset}, with input \texttt{x}, returns \texttt{TRUE} when \texttt{x} is of class \texttt{"hset"}, and \texttt{FALSE} otherwise. 
The function \texttt{as.hset}, with input \texttt{x}, return \texttt{x} itself if it is of class \texttt{"hset"}, otherwise it applies the constructor \texttt{hset}, with \texttt{members} equal to \texttt{x}, \texttt{multiplicities} and \texttt{generalized} as default, so that the function creates a set with elements taken from \texttt{x}.

Size and cardinality defined in equations (\ref{size}) and (\ref{cardinality}) are returned by the functions \texttt{size.support} and \texttt{cardinality}.
A vector containing the labels of the elements is returned by the function \texttt{members}, while the vector of multiplicities of the elements are returned by the function \texttt{multiplicities}.
The only input of these four functions is an object of class \texttt{"hset"}.
The components of the vectors obtained by the last two functions are coherent, so that the $i$-th value of the vector is the multiplicity of the $i$-th element.
If the function \texttt{multiplicities} is used on a set, a vector with all values equal to $1$ is returned.

\subsection{Semantic of \texttt{hset}}
\label{subsec: semantic of hset - sets and multisets} 
In R, objects can be accessed and modified with \emph{reference}, or with \emph{value semantic} (the two alternatives are described in Appendix \ref{sec: semantics in programming}).
In R objects of most classes have value semantic, but environments and hash tables from the package \texttt{hash} have reference semantics. 
An object of class \texttt{"hset"} contains a \texttt{hash} object in the first slot, and an \texttt{environment} object in the second. 
When a set is copied ``directly", both slots of the copied object, refer to the same ones of the original object.
The functions \texttt{clone.of.hset} and \texttt{refer.to.hset} are used to copy an \texttt{hset} with value and reference semantic, respectively.

The function \texttt{is.generalized}, with logical returned value, is used to distinguish sets and multisets.
The map \texttt{as.generalized} transforms a set to a multiset by converting the values of all elements of the hash table to $\texttt{1L}$, while \texttt{as.not.generalized} transforms a multisets to a set by converting all values to $\texttt{""}$.
These functions implement $X\mapsto Y(X)$ and $Y\mapsto X(Y)$ that were defined mathematically in Section \ref{subsec: mathematical definition - sets and multisets}.
The sets are modified locally, i.e., with reference semantic, so if a multiset is transformed to a set the information about the multiplicities of the multiset that is passed as input is lost.
The functions \texttt{clone.of.hset} and \texttt{refer.to.hset} can also be used to convert sets to multisets, and vice versa.
They have as second argument the (empty, or logical) value called \texttt{generalized}, that is \texttt{NULL} by default, but it can be used to convert a multiset to a set, or a set to a multiset, when it is copied.
Applying \texttt{refer.to.hset} with second argument equal to \texttt{TRUE} (respectively \texttt{FALSE}), is equivalent to applying the function \texttt{as.generalized} (respectively \texttt{as.not.generalized}).
Whereas the application of \texttt{clone.of.hset} with second argument equal to \texttt{TRUE} or \texttt{FALSE}, creates a new \texttt{hset} with the same support as the original one, but with the multiplicities that are converted to \texttt{1L} or \texttt{""} in the two cases, and the \texttt{hset} that is passed as input of \texttt{clone.of.hset} is not modified.

In the next section relations and operations between sets and multisets will be described mathematically and computationally.
In this section we describe how the chosen semantic can affect the computation of an operation.
Relations are encoded as functions with Boolean codomain, so if two components that can be elements, sets or multisets, are in relation, the function returns \texttt{TRUE}, otherwise \texttt{FALSE}.
As sets are not modified when checking whether two components are in relation, the semantic is irrelevant.
Conversely for operations between sets or multisets, even though the result of the operation is not changed by the semantic used, one operand will be modified when reference semantic is used, while a new \texttt{hset} with the result of the operation is created. 
The operands do not change when value semantic is used.

All operations are computed with the function \texttt{hset.operation.numeric} if at least one of the operands is a multiset, otherwise \texttt{hset.operation.logical} is used.
These two functions have the same signature in which the output is an object called \texttt{new.hset} of class \texttt{"hset"}. 
The first input \texttt{hset1} is the first operand, \texttt{...} contains all other operands (for operations with multiple arity), the arguments \texttt{operation} (function) and \texttt{identity.is.universe} (logical value) completely specify how the operands are combined. The last input \texttt{semantic}, that can be equal to \texttt{"refer"} (default) or \texttt{"value"}, specifies the semantic.
The difference between the numeric and the logical operation is in how the multiplicities are handled.
In the latter case we define the multiplicities of a set by the bijection that maps \texttt{NULL} to \texttt{FALSE} and \texttt{""} to \texttt{TRUE}, where the Boolean outcomes are used to evaluate the operation.
The evaluation is stored using the inverse function, as setting an element to \texttt{NULL} in an \texttt{hash} object is equivalent to removing the key-value pair from the hash table, or to not doing anything if the pair is not present.
If the numeric operation is used, the multiplicities that are used in the operation are obtained by the surjective function of type $\texttt{NULL}\uplus \texttt{""}\uplus  \texttt{N}_{\texttt{+}}\to \texttt{N}_{\texttt{+=}}$, s.t. $\texttt{NULL}\mapsto \texttt{0}$, $\texttt{""}\mapsto \texttt{1}$, and $\texttt{m}\mapsto \texttt{m}$ for $\texttt{m}\in \texttt{N}_{\texttt{+}}$.
The non-negative values are then used as operands, and the result is stored in the hash table with the bijection $\texttt{N}_{\texttt{+=}}\to \texttt{NULL}\uplus \texttt{N}_{\texttt{+}}$, s.t. $\texttt{0}\mapsto \texttt{NULL}$, and $\texttt{m}\mapsto \texttt{m}$ for all $\texttt{m}\in \texttt{N}_{\texttt{+}}$.
Note that the output of a numeric operation is always a multiset, so \texttt{""} is never stored.

The function \texttt{create.new.hset} is used in \texttt{hset.operation.numeric} and \texttt{hset.operation.logical}, to create the object \texttt{new.hset} that will store the result.
When reference or value semantics are used, \texttt{refer.to.hset} or \texttt{clone.of.hset} are used respectively, inside \texttt{create.new.hset}, with argument \texttt{hset1}.
Therefore, \texttt{new.hset} and \texttt{hset1} will refer to the same object in memory with reference semantic, so that when the result is computed, will be stored both in \texttt{new.hset}, and in \texttt{hset1}.
Whereas with value semantic, \texttt{new.hset} will be a reference to an object in memory that is a clone of \texttt{hset1}, so the latter will not change when the result is computed.
The reference semantic is used by default for its computational advantages. 
In particular when the identity element of the operation is the empty set, that is when \texttt{identity.is.universe} is \texttt{FALSE}, computing the result does not require a complete scan through the elements of \texttt{hset1}.

The computational complexity of an operation between multisets $Y_1$, $Y_2$, $...$, $Y_a$, where $\emptyset$ is the identity element of the operation, is $O(|Y_2|+...+|Y_a|)$ with reference semantic, and $O(|Y_1|+|Y_2|+...+|Y_a|)$ with value semantic.
The advantage is significant when $|Y_1|\gg \max_{j\ne 1}|Y_j|$.
Note that difference of $O(|Y_1|)$ operations between the two semantics, is due to the necessity of copying the first operand.
However, if the first operand is a set, while some of the others are multisets, there is no difference between the two semantics, because the result of the operation is a multiset, so $O(|Y_1|)$ operations are required to convert the first operand to a multiset, even when reference semantic is used.

\section{Sets algebra}
\label{sec:sets algebra}

Relations and operations involving \texttt{hset} objects are described in Sections \ref{subsec:relations - sets algebra} and \ref{subsec:operations}, respectively.
Relations of different types are described by their signature, definition, and implementation as functions with Boolean codomain.
Operations are also described in the same way, but the functions that compute them return sets or multisets. 

\subsection{Relations}
\label{subsec:relations - sets algebra}

\paragraph{Inclusion of elements.} The \emph{inclusion} relation between an element and a set is defined mathematically as
\begin{equation}
\in :(\texttt{N}\uplus \texttt{S})\times \texttt{S}, \ \ \ \    a \in X \ \iff \ X=\{a,...\},
\label{inclusion.of.element.in.set}
\end{equation}
meaning that the element $a\in (\texttt{N}\uplus \texttt{S})$ and the set $X \in \texttt{S}$ are related by $\in$ if and only if $a$ is an element of $X$.
Note that $\emptyset\in X$ for all sets $X$, and that the symbol $:$ is used in the signature of the relation to avoid the notation $\in\in(\texttt{N}\uplus \texttt{S})\times \texttt{S}$.
This relation is extended trivially to multisets as
\begin{equation}
\in :(\texttt{N}\uplus \texttt{S})\times \texttt{M}, \ \ \ \
a \in Y \ \iff \ Y=\{a[n],...\}, \ n\ge 1,
\label{inclusion.of.element.in.multiset}
\end{equation}
that is if the multiplicity of $a\in (\texttt{N}\uplus \texttt{S})$ in $Y \in \texttt{M}$ is at least $1$.

The straightforward generalization for multisets are relations between an element with a given multiplicity, and a multiset.
Three relations of this type are defined as 
\begin{equation}
\in_{\sim} :((\texttt{N}\uplus \texttt{S})\times \texttt{N}_{\texttt{+}}) \times \texttt{M}, \ \ \ \
    a[m] \in_\sim Y \ \iff \ Y=\{a[n],...\}, \ m-n\sim 0,    
\end{equation}
where $\sim$ can be $\le$, $<$ and $=$, for the relations of, inclusion $\in = \in_{\le}$, \emph{strict inclusion} $\in_{<}$ and \emph{exact inclusion} $\in_{=}$, respectively.
Intuitively, in the three cases, $a[m]$ and $Y$ are related if $a$ in $Y$ has a multiplicity greater or equal, greater, and equal to $m$ respectively.
Instead of defining relations between an element with a given multiplicity and a multiset, we could have equivalently defined the family of relations between elements and multisets parametrized by $m\in \texttt{N}_{\texttt{+}}$:
\begin{equation}
\in_{\sim}^m :(\texttt{N}\uplus \texttt{S})\times \texttt{M}, \ \ \ \
    a \in_\sim^m Y \ \iff \ Y=\{a[n],...\}, \ m-n\sim 0,    
\end{equation}
so that, for all $a\in \texttt{N}\uplus \texttt{S}$, $Y\in \texttt{M}$, and $m\in\texttt{N}_{\texttt{+}}$,
\begin{equation}
a[m] \in_\sim Y \ \iff \ a \in_\sim^m Y. \end{equation}

All relations defined above can be encoded as one function with signature
\begin{equation}
((\texttt{N}\uplus \texttt{S})\uplus((\texttt{N}\uplus \texttt{S})\times \texttt{N}_{\texttt{+}})) \times (\texttt{S} \uplus \texttt{M}) \times \{\le,<,=\}\to \{\texttt{TRUE},\texttt{FALSE}\},    
\end{equation}
where the first argument is either an element, or a pair between an element and a multiplicity, the second argument is either a set or a multiset, whereas the last argument specifies the type of relation.
The function returns \texttt{TRUE} if the first two arguments are in relation, of the type specified by the third argument.
In the package, a similar function, called \texttt{inclusion.member}, has signature
\begin{equation}
\texttt{C} \times (\texttt{S} \uplus \texttt{M}) \times \texttt{N}_{\texttt{+}} \times \{\le,<,=\}\to \{\texttt{TRUE},\texttt{FALSE}\}.   
\end{equation}
The last two arguments, called \texttt{multiplicity} and \texttt{type.relation} with default values \texttt{1} and $\le$ respectively, are ignored when the second argument is a set.
The first argument, called \texttt{member} must be a vector of length 1, that is converted to a character \texttt{C} inside the function, if this value is not a valid element, as defined above, the function returns \texttt{FALSE} for every possible choice of the last two arguments.
Then, \texttt{inclusion.member} returns \texttt{TRUE} if and only if the first two arguments are in relation specified by the last two arguments.
The binary operator \texttt{\%in\%} uses \texttt{inclusion.member} where the last two arguments are set as default, so that it evaluates the relations (\ref{inclusion.of.element.in.set}) or (\ref{inclusion.of.element.in.multiset}), depending on whether the second argument is a set or a multiset respectively.
If the first argument, i.e., the left operand of \texttt{\%in\%}, is a vector of characters, the operand returns a vector of Booleans with the result of the evaluated relation for each character. 

\paragraph{Subsets and equalities.} Now relations in which both objects are sets or multisets are considered.
The \emph{subset relation} between sets $X_1$ and $X_2$ is
\begin{equation}
\subseteq:\texttt{S}\times \texttt{S}, \ \ \ \ X_1\subseteq X_2 \iff (a\in X_1 \implies a\in X_2),
\end{equation}
and the \emph{strict subset relation} is
\begin{equation}
\begin{split}
\subset:\texttt{S}\times \texttt{S}, \ \ \ \ X_1\subset X_2 &\iff (X_1\subseteq X_2 \ \mathrm{and} \ \exists b\in X_2 \ \mathrm{s.t.} \ b\notin X_1).
\end{split}
\end{equation}
The \emph{equality relation} between sets is defined as
\begin{equation}
\begin{split}
=:\texttt{S}\times \texttt{S}, \ \ \ \ X_1= X_2 &\iff (X_1\subseteq X_2 \ \mathrm{and} \ X_2\subseteq X_1).
\end{split}
\end{equation}
For finite dimensional sets, the last two relations can also be written as
\begin{equation}
    X_1 \approx X_2 \iff X_1 \subseteq X_2 \ \mathrm{and} \ |X_1| \sim |X_2|,
\end{equation}
where for the strict inclusion, $\approx$ and $\sim$ are replaced by $\subset$ and $<$ respectively, while for the equality relation $\approx$ and $\sim$ are both replaced by $=$.

The relations above will be generalized in the case in which at least one component of the relation is a multiset.
For $Y=\{a_i[m_i]\}\in \texttt{M}$, the multiplicities are written as a function $v_Y:X(Y) \to \texttt{N}_{\texttt{+}}$, such that $v_Y(a_i)=m_i$.
The domain of this function, that is the support of $Y$, is extended to all well defined elements, as
\begin{equation}
v_Y:\texttt{S}\uplus \texttt{N} \to \texttt{N}_{\texttt{+}}\cup\{0\}, \ \ \text{s.t.} \ \ v_Y(a_i)=
\begin{cases}
m_i &\text{if} \ a_i\in X(Y)\\
0 &\text{otherwise}
\end{cases}.   
\label{generalized.multiplicity.containing.zeros}
\end{equation}
Note that the extension of this function is coherent with equation (\ref{multiset.with.element.with.zero.multiplicity}).
The relations above are generalized as
\begin{equation}
\begin{split}
&Y_1 \subseteq Y_2 
\iff v_{Y_1}(a) \le v_{Y_2}(a) \ \forall a\in \texttt{S}\uplus \texttt{N}
, \\
&Y_1 \subset Y_2 \iff Y_1 \subseteq Y_2 \ \mathrm{and} \ \exists a\in \texttt{S}\uplus \texttt{N} \ \mathrm{s.t.} \ v_{Y_1}(a) < v_{Y_2}(a), \\
&Y_1 \sqsubseteq Y_2 \iff Y_1 \subseteq Y_2 \ \mathrm{and} \ \nexists a\in \texttt{S}\uplus \texttt{N} \ \mathrm{s.t.} \ 0\ne v_{Y_1}(a) < v_{Y_2}(a),\\
&Y_1 \sqsubset Y_2 \iff Y_1 \sqsubseteq Y_2 \ \mathrm{and} \ \exists a\in \texttt{S}\uplus \texttt{N} \ \mathrm{s.t.} \ 0=v_{Y_1}(a) < v_{Y_2}(a), \\
&Y_1 = Y_2 \iff Y_1 \subseteq Y_2 \ \mathrm{and} \ Y_2 \subseteq Y_1 \iff v_{Y_1}(a) = v_{Y_2}(a) \ \forall a\in \texttt{S}\uplus \texttt{N},
\end{split}
\label{subset.relations.for.multisets.definition}
\end{equation}
The signature of these relation is $\approx: \texttt{S}\uplus \texttt{M} \times \texttt{S}\uplus \texttt{M}$, where $\approx$ is one of the five relations above.
However, only the definition for multisets is given in (\ref{subset.relations.for.multisets.definition}), but this is not a problem, as equation (\ref{conversion.set.to.multiset.Y(X)}) implies that if at least one argument, say the first one, is a set, the relation $X_1\approx Y_2$ is equivalent to $Y(X_1)\approx Y_2$.

For finite dimensional multisets, the relations can be written as 
\begin{equation}
\begin{split}
&Y_1 \subseteq Y_2 
\iff  v_{Y_1}(a) \le v_{Y_2}(a) \ \forall a\in X(Y_1)
, \\
&Y_1 \subset Y_2 \iff Y_1 \subseteq Y_2 \ \mathrm{and} \ (\exists a\in X(Y_1) \ v_{Y_1}(a) < v_{Y_2}(a) \ \mathrm{or} \ |Y_1|<|Y_2|), \\
&Y_1 \sqsubseteq Y_2 \iff v_{Y_1}(a) = v_{Y_2}(a) \ \forall a\in X(Y_1) \ \mathrm{and} \ |Y_1|\le|Y_2|,\\
&Y_1 \sqsubset Y_2 \iff v_{Y_1}(a) = v_{Y_2}(a) \ \forall a\in X(Y_1) \ \mathrm{and} \ |Y_1|<|Y_2|,\\
&Y_1 = Y_2 \iff v_{Y_1}(a) = v_{Y_2}(a) \ \forall a\in X(Y_1) \ \mathrm{and} \ |Y_1|=|Y_2|.
\end{split} 
\label{subset.relations.for.multisets.definition.2}
\end{equation}
Other definitions are available, but these ones are the most efficient computationally, because it is not necessary to evaluates multiplicities in $Y_2$ for elements not in $X(Y_1)$. 
The function \texttt{hset1.included.to.hset2} with signature
\begin{equation}
(\texttt{S}\uplus\texttt{M})\times (\texttt{S}\uplus\texttt{M})\times \{\texttt{TRUE},\texttt{FALSE}\}\times \{\texttt{TRUE},\texttt{FALSE}\}\to \{\texttt{TRUE},\texttt{FALSE}\}    
\end{equation}
where the four arguments are \texttt{hset1}, \texttt{hset2}, \texttt{strictly} and \texttt{exactly}, returns \texttt{TRUE} if the first two arguments are in one of the relations defined above, except for the equality, that is implemented with another function.
If the first two arguments are both sets, the fourth argument is ignored because in $\texttt{S}\times \texttt{S}$ the relations $\subseteq$ and $\sqsubseteq$ are equivalent, and so are $\subset$ and $\sqsubset$.
The function iterates through the members of \texttt{hset1}, computes the difference of multiplicities, if this difference is negative \texttt{FALSE} is returned immediately, otherwise the difference is accumulated.
For $\sqsubseteq$ and $\sqsubset$ the accumulated difference must be zero at the end of the loop, and the two relations are distinguished by the condition on the supports.
For $\subset$, either the accumulated difference is strictly positive, or the support of the second set is larger.
Whereas for $\subseteq$, no other conditions are required after the end of the loop.
The evaluation of the equality relations is implemented in the function \texttt{hsets.are.equal}, in which only the functions \texttt{size.support}, \texttt{members} and \texttt{multiplicities}
defined at the end of Section \ref{subsec: mathematical definition - sets and multisets} are used to evaluate the relation.

Some generic operators that call the function \texttt{hset1.included.to.hset2} with different combinations of the last two inputs are defined.
For $\subseteq$, \texttt{<=} and \texttt{>=} are used, where the latter is for the inverse relation $\supseteq$, obtained by reflecting the arguments.
For $\subset$, the generic operators are \texttt{<} and \texttt{>}, for $\sqsubseteq$ they are \texttt{\%=<=\%} and \texttt{\%=>=\%}, and for $\sqsubset$, \texttt{\%=<\%} and \texttt{\%=>\%} are used.
The equality operator \texttt{==} calls the function \texttt{hsets.are.equal} and \texttt{!=} returns its negation.

\subsection{Operations}
\label{subsec:operations}
When discussing relations the semantic of the implementation was never mentioned, as the sets that were possibly be part of some relations were never modified, and returned by the functions used to check whether two objects are in relation.
An \emph{operation} is a ternary relation between sets or multisets, that is written as $X_1\approx X_2 = X$ for a given $\approx$, however the interest here is computing $X$ from the \emph{operands} $X_1$ and $X_2$.
Moreover all operations will be defined for general arity, that is for more than two operands.
The \emph{universe set}, denoted by $U$ is a set such that $X\subseteq U$, for all $X\in\texttt{S}$, that is the set containing all possible elements: $a\in U$ for all $a\in \texttt{S}\uplus\texttt{N}$. 
The \emph{universe multiset} $U$ is the multiset such that $Y\subseteq U$, for all $Y\in\texttt{M}$.
Then, $U$ contains all elements in $\texttt{S}\uplus\texttt{N}$, each with infinite multiplicity.
Note that the universe set and multiset are defined by the same symbol, as it will be clear from the context which ``universe" is considered.

The signature of all operations defined below is
\begin{equation}
\approx:(\texttt{S}\uplus\texttt{M})\times \biguplus_{k\ge 0}(\texttt{S}\uplus\texttt{M})^k \to (\texttt{S}\uplus\texttt{M}),
\end{equation}
so that the operation is defined if there is at least one operand, the second argument is written as a disjoint union of all possible tuple of \texttt{hset}s, so that for a given $k$, there are $k+1$ operands.
However, only binary operations are defined, as the extension to general arities is straightforward.
The result of the operation is of class $\texttt{S}$ if and only if all operands are of class $\texttt{S}$, otherwise the result is of class $\texttt{M}$.
In the latter case, if an operand $X_i$ is a set, it is replaced by $Y_i=Y(X_i)$.

The \emph{intersection}, \emph{union}, \emph{sum}, \emph{difference} and \emph{symmetric difference} between $X_1$ and $X_2$ are
\begin{equation}
\begin{split}
&X = X_1 \cap X_2 \iff (a\in X\implies a\in X_1 \ \text{and} \ a\in X_2), \\
&X = X_1 \cup X_2 \iff (a\in X\implies a\in X_1 \ \text{or} \ a\in X_2), \\
&X = X_1 + X_2 \iff X = X_1 \cup X_2,\\
&X = X_1 \setminus X_2 \iff (a\in X\implies a\in X_1 \ \text{and} \ a\notin X_2), \\
&X = X_1 \triangle X_2 \iff X = (X_1 \setminus X_2) \cup (X_2 \setminus X_1),
\end{split}
\end{equation}
respectively.
The identity element for the intersection is $X_2=U$, while for all other operations is $X_2=\emptyset$.
All operations, except for the difference, are commutative and associative, $X_1\setminus X_2\setminus X_3$ is defined to be equal to $(X_1\setminus X_2)\setminus X_3$.
Note that the sum has been defined to be the same as the union, but these operations will be different for multisets.
The multiset version of the operations above is
\begin{equation}
\begin{split}
&Y = Y_1 \cap Y_2 \iff \forall a, v_Y(a) = \min(v_{Y_1}(a), v_{Y_2}(a)), \\
&Y = Y_1 \cup Y_2 \iff \forall a, v_Y(a) = \max(v_{Y_1}(a), v_{Y_2}(a)), \\
&Y = Y_1 + Y_2 \iff \forall a, v_Y(a) = v_{Y_1}(a) + v_{Y_2}(a),\\
&Y = Y_1 \setminus Y_2 \iff \forall a, v_Y(a) = \max(v_{Y_1}(a) - v_{Y_2}(a),0), \\
&Y = Y_1 \triangle Y_2 \iff \forall a, v_Y(a) = |v_{Y_1}(a) - v_{Y_2}(a)|.
\end{split}
\end{equation}
As for the set version, the identity elements are $Y_2=U$ and $Y_2=\emptyset$ for the intersection, and for all other operations respectively.
When generalizing $\setminus$ and $\triangle$ to multisets, some properties that hold for sets are violated.
For example, in general $(Y_1\setminus Y_2)\cap Y_2\ne\emptyset$, and $Y_1\triangle Y_2\triangle Y_3 \ne Y_2\triangle Y_1\triangle Y_3$.

The functions \texttt{hset.operation.numeric} and \texttt{hset.operation.logical} have signature
\begin{equation}
\begin{split}
&(\texttt{S}\uplus\texttt{M})\times (\texttt{S}\uplus\texttt{M})^* 
\times  ((\texttt{N}_\texttt{+}\uplus \texttt{0})^*\to \texttt{N}_\texttt{+}\uplus \texttt{0})\times
\texttt{B}\times \{\texttt{"refer"},\texttt{"value"}\} \to \texttt{M},\\
&\texttt{S}\times \texttt{S}^* 
\times  (\texttt{B}^{*}\to \texttt{B})\times
\texttt{B}\times \{\texttt{"refer"},\texttt{"value"}\} \to \texttt{S},
\end{split}
\end{equation}
respectively, where $\texttt{Z}^*$ is computed as in equation (\ref{eq:sequences}), and $\texttt{B}=\{\texttt{FALSE},\texttt{TRUE}\}$.
The first two arguments, called \texttt{hset1} and \texttt{...}, contain the operands.
The third argument, called \texttt{operation}, is a function that computes the updated multiplicity, the choice of this function defines which of the operations above (intersection, union, sum, difference, symmetric difference) is computed. 
The fourth argument called \texttt{identity.is.universe}, with default \texttt{FALSE}, specifies whether iterating through all elements of the first operand is needed.
The last argument called \texttt{semantic}, with default \texttt{"refer"}, specifies whether the first operand is modified, or whether its clone is modified.
For the intersection, union, sum, difference, and symmetric difference, with logical multiplicities, the functions that are used in the third input are \texttt{all}, \texttt{any}, \texttt{any}, \texttt{nimp} (for ``not implies") and \texttt{niff} (for ``not if and only if") respectively.
Whereas, with numeric multiplicities, the functions are \texttt{min}, \texttt{max}, \texttt{sum}, \texttt{pdif} (for ``positive difference") and \texttt{sdif} (for ``symmetric difference") respectively.
The functions \texttt{nimp}, \texttt{niff}, \texttt{pdif} and \texttt{sdif} have been implemented by us, while the others are primitive functions in R.
In the intersection the fourth argument is \texttt{TRUE} ($U$ is the identity element of the operation), while in all other operations, the fourth argument is \texttt{FALSE} ($\emptyset$ is the identity element).

The functions \texttt{intersection}, \texttt{union}, \texttt{setsum}, \texttt{difference}, \texttt{symmdiff} with signature 
\begin{equation}
    (\texttt{S}\uplus\texttt{M})\times (\texttt{S}\uplus\texttt{M})^* \times  \{\texttt{"refer"},\texttt{"value"}\} \to (\texttt{S}\uplus\texttt{M}),
\end{equation}
call \texttt{hset.operation.numeric} if at least one operand is a multiset, otherwise \texttt{hset.operation.logical} is used, with third and fourth arguments set by the operation.
The infix operands for computing the binary intersection are \texttt{\%\&\%}, \texttt{\%\&\&\%} and \texttt{\%and\%}, for the union, the operands are \texttt{\%|\%}, \texttt{\%||\%} and \texttt{\%or\%}, for the sum are \texttt{\%+\%} and \texttt{\%sum\%}, for the difference \texttt{\%-\%} and \texttt{\%!implies\%}, and for the symmetric difference \texttt{\%--\%} and \texttt{\%xor\%}.
In all these operands reference semantic is used, for a binary operation with value semantic, all operands can be used, but a $\sim$ is added before the last $\texttt{\%}$, e.g., for the union with value semantic, \texttt{\%|}$\sim$\texttt{\%} can be used.

\section{Performance}
\label{sec:performance}
Here we assess the performance of our implementation, by comparing its two semantics between themselves, and with the package \texttt{sets} \citep{sets}.

\subsection{Relations}
Computing a relation results to a Boolean output indicating whether the two components are related. 
Therefore the sets are not modified and the semantic is irrelevant.
The only comparison is then between our implementation based on an hash table, denoted here by \texttt{hsets}, and the implementation of the library \texttt{sets}.

\begin{figure}
    \centering
    \includegraphics[scale=.53]{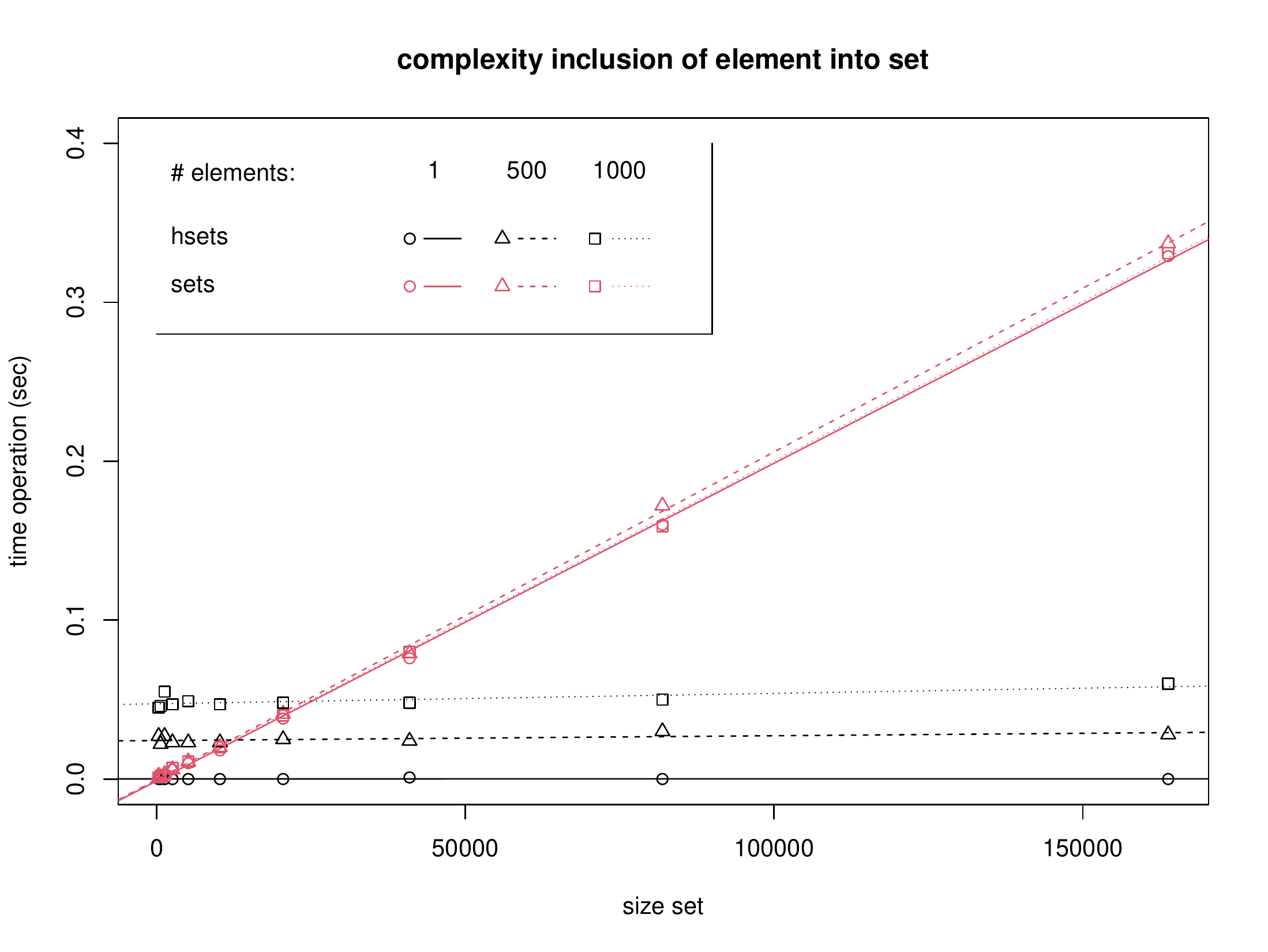}
    \caption{Time complexity for evaluating the inclusion relation.}
    \label{fig:relations}
\end{figure}

The comparison for the inclusion relation $a \in X$ from equation (\ref{inclusion.of.element.in.set}) between the element $a$ and the set $X$, is in Figure \ref{fig:relations}.
The \emph{x} axis is $|X|$, which is the size of the set $X$, on the \emph{y} axis the time of evaluating the relation $\texttt{ai \%in\% X}$, where $\texttt{ai}$ is a vector of elements, and $\texttt{X}$ have classes $\texttt{hset}$ and $\texttt{sets}$, for the black and the red dots, respectively.
The shape of the dots denotes for how many elements, that is the size of $\texttt{ai}$, the relation is evaluated.
In our implementation, the complexity of the operation does not depend on the size of the set $|X|$, but it depends linearly on the size of $\texttt{ai}$, that is the number of elements that we want to check whether they are contained in the set.
On the other hand, the complexity of the implementation of $\texttt{sets}$, depends linearly on $|X|$, but it seems not to depend on the size of $\texttt{ai}$, implying that the inclusion relations onto a set are parallellized in $\texttt{sets}$.

In our implementation, the subset and the equality relations between sets, written succinctly as $X_1\approx X_2$, with $\approx$ in $\{\subseteq, \subset, =\}$, have the same complexity evaluated above, that is constant with respect to $|X_2|$, and linear with respect to $|X_1|$, where $X_2=X$ and $X_1=\{a_i\}_i$.
The same complexity can be found if $X_1$ and $X_2$ are replaced by multisets $Y_1$ and $Y_2$, for all cases of $\approx$.
The reason is that, when evaluated using the formulas in (\ref{subset.relations.for.multisets.definition.2}), the complexity of the relation grows linearly with respect to $|Y_1|$ because of the component $\forall a\in X(Y_1)$, and the complexity is constant with respect to $|Y_2|$ because $|Y_1| \approx |Y_2|$ is evaluated in constant time, for various $\approx$.

\subsection{Operations}
\label{sec:operationsComplexity}
\begin{figure}
    \centering
    \includegraphics[scale=.59]{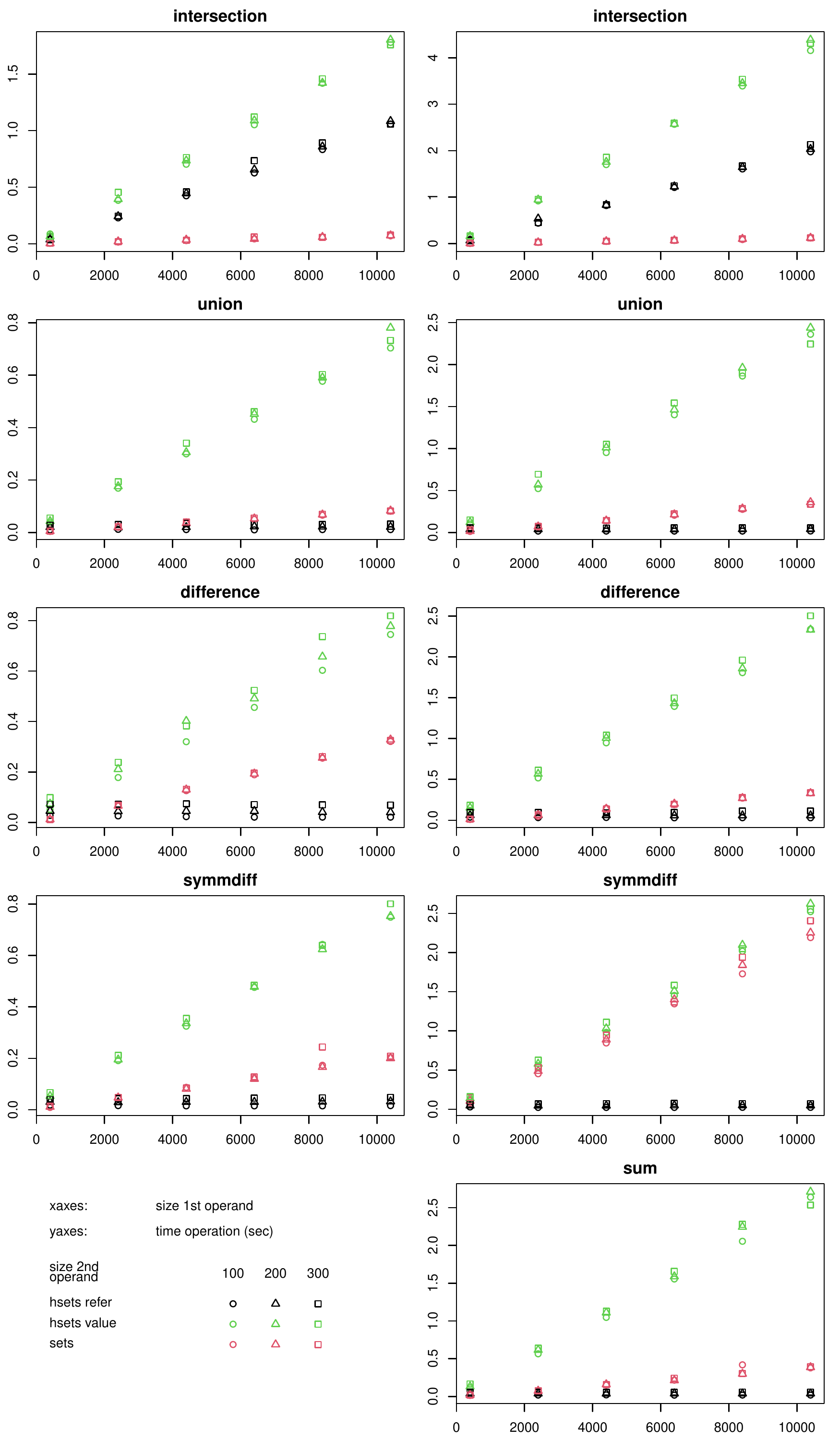}
    \caption{Time complexity of set (left) and multiset (right) operations for different implementations, and sizes of the operands.}
    \label{fig:operations}
\end{figure}

The codomain of an operation is either a set or a multiset.
In Figure \ref{fig:operations} the complexity of the defined set operations is plotted in the left column.
The \emph{x} and \emph{y} axes distinguish the size of the operands and the time for evaluating one operation in seconds.
The shape of the points denotes the size of the second operands, their colours distinguish the semantics of our implementation and the operations computed with objects from the package \texttt{sets}.
In Section \ref{subsec: semantic of hset - sets and multisets}, it has been explained that the reference semantic is helpful for some operations when the first operand is large.
In particular, the time complexity of the operation is constant with respect to the size of the first operand, when $\emptyset$ is the identity element of the operation, that is for the union, difference and symmetric difference.
The complexity cost of using the value semantic can be seen by the fact that the complexity grows linearly with the size of both operands, as the (large) first one has to be copied at the beginning of the operation.
Whereas with the reference semantic the complexity of these operation is linear only with respect to the size of the second operand, while being constant with respect to the size of the first one.
In the library \texttt{sets}, the complexity of the operations grows linearly with respect to the size of the first operand, but it seems not to depend on the size of the second one.
We think that it actually depends on the size of the largest operands.
For the interaction however our implementation is 
much more inefficient that \texttt{sets}, regardless of the semantic, although all implementations have a linear complexity with respect to the size of the first operand.
The sum of two sets is defined to be equivalent to the union, so the complexity of the implementations is not computed.

In the right column of Figure \ref{fig:operations} the same comparison has been done for operations between multisets.
As in the previous case, the intersection is much more efficient in the \texttt{sets} package, whereas for all other operations, our implementation with reference semantic does not depend on the size of the first operand, while with value semantic it does, as the first operand is copied.

\section{MCMC with state space of undirected graphs}
\label{sec:application}

In this section we describe an application of the \texttt{hset} package to \emph{Markov processes} with set- or multiset-valued (discrete) \emph{state space}.
The example considered here is a stochastic process with set-valued state space, and multiset-valued \emph{sufficient statistic} for the distribution of the process.
The stochastic process is defined over the $n(n-1)/2$ dimensional state space of $n$ dimensional \emph{undirected simple graphs}. 
Each graph is represented uniquely by the \emph{edge set}, which contains all information (ties) about the graph.
The state of the process is augmented with the \emph{degree distribution} of the graph, that is a multiset and contains all relevant information about the distribution of the stochastic process.

The network process evolves by \emph{tie flips} in which (typically few) non-edges become edges and vice versa.
The tie flips are often \emph{local}, e.g., when only one tie can be flipped at given time, in general at each time point only a number of ties that is small, in comparison with the size of the network, can be flipped.
If few edges of the graph are flipped, the sufficient statistic does not have to be recomputed from the edge set, as only the degrees of the vertices adjacent to the flipped edges must be updated.
Therefore the reference semantic in our implementation that is derived from the hash table data structure, has the computational advantage of being able to modify the state locally (in memory), without having to copy the state when its size changes.


We consider here the \emph{Beta Model} \citep{holland1981exponential, blitzstein2011sequential, MLEbetaRinaldo}, where vertex $i$ has its own parameter $\beta_i$, and the tie $ij$ between vertices $i$ and $j$ is in the graph with probability $p_{ij}=e^{\beta_i+\beta_j}/(1+e^{\beta_i+\beta_j}).$
The model is equivalent to a sample of $n(n-1)/2$ independent Bernoulli random variables $X_{ij}$ with probabilities $p_{ij}$.
The distribution of the network can be written as 
\begin{equation}
P(X=x|\beta) = \prod_{1\le i< j\le n} p_{ij} = \exp\Big(\sum_{i=1}^n d_i(x)\beta_i - \psi(\beta)\Big),
\label{eq:probabilityDensityBetaModel}
\end{equation}
where $d_i(x)$ is the degree of vertex $i$, and $\psi(\beta)=\sum_{i<j}\log(1+e^{\beta_i+\beta_j})$ is the log-partition function.
Thus the \emph{degree vector} $d=(d_1,...,d_n)$ is a sufficient statistic for the model.

We consider a Markov chain in which at each time point some (typically few, in comparison with $n$) tie flips are proposed, and accepted with probability 
\begin{equation}
Q(\tilde{x}|x,\beta) = \min\Big(\exp\Big({\sum}_{i\in \mathcal{I}}(\tilde{d}_i-d_i)\beta_i\Big),1\Big),
\label{eq:acceptanceProbabilityMCMCbetaModel}
\end{equation}
where $\mathcal{I}\subseteq\{1,...,n\}$ and $i\in I$ if exist vertex $j$ such that the tie $ij$ is included in the proposed tie flips (thus $i\in \mathcal{I}\iff j\in \mathcal{I}$), $d_i = d_i(x)$ and $\tilde{d}_i = d_i(\tilde{x})$ are the degrees of vertex $i$ before and after the tie flips, respectively.
If the proposed tie flip contains only the tie $ij$, then $\mathcal{I}=\{i,j\}$ and $|\mathcal{I}|=2$, in general more than one tie can be flipped, but the algorithm is useful computationally when $|\mathcal{I}|\ll n$.
Note that this process is a Markov chain with \emph{Metropolis} updates, as we use a symmetric proposal distribution for the tie flips (uniform distribution over $n(n-1)/2$).
Therefore, the stationary distribution of the Markov chain with acceptance probability $Q(\tilde{x}|x,\beta)$ in equation (\ref{eq:acceptanceProbabilityMCMCbetaModel}) is $P(X=x|\beta)$ in equation (\ref{eq:probabilityDensityBetaModel}).

Three Markov chains $(X_t)_t$, $({X_t}^-)_t$ and $({X_t}^+)_t$ are considered, with same transition probability parametrized by $\beta$, but with different starting points.
The chains $(X_t)_t$, $({X_t}^-)_t$ and $({X_t}^+)_t$ have \emph{stationary}, \emph{sparse} and \emph{dense starting} point, respectively.
Therefore $X_0$, ${X_0}^-$ and ${X_0}^+$ have degrees similar, lower and higher, respectively, than the expected degrees computed from the stationary distribution of the Markov chain.
The real parameter is generated as $\beta\sim \mathrm{Norm}(-1_n,I_n)$. 
At each iteration, a single tie flip is proposed, sampling $h\sim \mathrm{Unif}(1,...,n(n-1)/2)$, and the tie $ij$ is computed from $h$.

With our implementation, the state of the chain with stationary starting point is $(X_t, Z_t)$, that is composed by two objects of class \texttt{"hset"}, $X_t$ is the network encoded as set of edges, and $Z_t$ is the degree distribution of the network encoded as a multiset.
In each iteration the \texttt{hset} of flips $F$ is sampled, in our case $|F|=1$ as only one tie is sampled.
Then the set $\mathcal{I}_F$ of vertices that are part of at least one proposed tie flip is computed, and so are the proposed degrees $\tilde{d}_i$ for $i\in \mathcal{I}_F$.
If the flip is accepted, the state is updated as
\begin{equation}
\begin{split}
&X_{t+1} \xleftarrow{} X_{t} \ \triangle \ F, \\
&Z_{t+1} \xleftarrow{} (Z_{t} - \{d_i[m_i]: i\in \mathcal{I}_F\}) + \{\tilde{d}_i[\tilde{m}_{i}]: i\in \mathcal{I}_F\},
\end{split}
\label{eq: updates Xt Zt}
\end{equation}
where $\tilde{m}_i$ and $m_i$ are the multiplicities of proposed $\tilde{d}_i$ and current $d_i$ degrees respectively, for $i\in\mathcal{I}_F$.
The update uses the operations $\triangle: \texttt{S}\times\texttt{S}\to\texttt{S}$, $+: \texttt{M}\times\texttt{M}\to\texttt{M}$ and $-: \texttt{M}\times\texttt{M}\to\texttt{M}$, that have been defined in Section \ref{subsec:operations}.
In Section \ref{sec:operationsComplexity} it has been shown that for these three operations (that all have the empty set / multiset as the identity second operand), the complexity is not affected by the sizes of $X_t$ and $Z_t$, but it depends linearly on the sizes of $F$ and $\mathcal{I}_F$.
The same algorithm is used with $({X_t}^+, {Z_t}^+)$ and $({X_t}^-, {Z_t}^-)$, in which the starting point is out of equilibrium .

The updates in equation (\ref{eq: updates Xt Zt}) are coded with reference semantic as 
\begin{equation}
\begin{split}
&\texttt{state\$edge.set \%xor\% hset(flips\$id)}
\\
&\texttt{state\$degree.frequencies \%-\% hset(names(table.old.degrees),}\\ &\texttt{  
   as.integer(table.old.degrees))}\\
&\texttt{state\$degree.frequencies \%+\% hset(names(table.new.degrees),}\\ &\texttt{  
  as.integer(table.new.degrees))}\\
\end{split}
\end{equation}
where \texttt{state\$edge.set} and \texttt{state\$degree.frequencies} are \texttt{hset} objects containing the current network and degree distribution, \texttt{flips\$id} contains the ties that are flipped, \texttt{table.old.degrees} and \texttt{table.new.degrees} contain the old and new degrees for the vertices in $\mathcal{I_F}$.
Note that the constructor \texttt{hset} is used to create the set $F$, and the multisets containing the degree frequencies to be subtracted and added. 

\begin{figure}
    \centering
    \includegraphics[scale=.36]{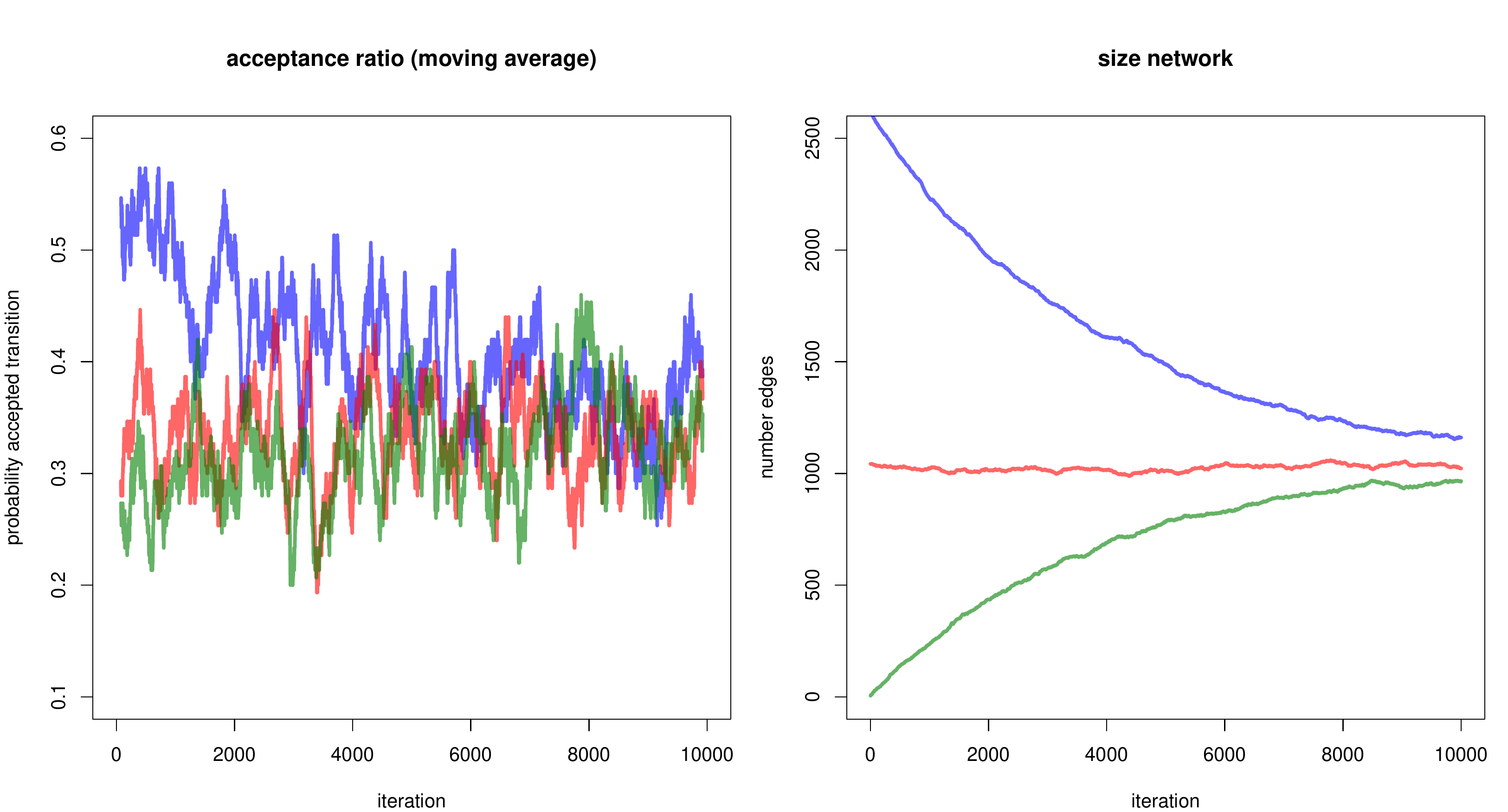}
    \caption{Left: acceptance ratio (moving average with $150$ filtered observations) of the Metropolis-Hastings algorithm. Right: number of ties of the network (state) at different iterations.
    Colours distinguish the different starting points: red for stationary, green for sparse, blue for dense.}
    \label{fig:MHiterations}
\end{figure}

In Figure \ref{fig:MHiterations} the processes derived from $(X_t)_t$, $({X_t}^-)_t$ and $({X_t}^+)_t$ are plotted in red, green and blue respectively.
In the left, the moving average of the acceptance ratio is plotted. 
In each iteration the proposed transition is either accepted, or it is not.
This binary outcome is replaced in the plot by the average of 150 binary values around it, giving an indication on how probable are transitions of state in and out of equilibrium.
The stationary and the sparse chain have a similar behaviour with transition probability approximately equal to $0.35$ throughout their whole history. 
The dense chain starts with a larger transition probability, that is reduced toward the equilibrium value as the process approaches the stationary distribution.
In the right plot the size of the state, that is the number of ties in the network is plotted in the three cases, showing how the distribution of $|{X_t}^-|$ and $|{X_t}^+|$ approach the one of $|{X_t}|$ as $t\to\infty$.

\begin{figure}
    \centering
    \includegraphics[scale=.36]{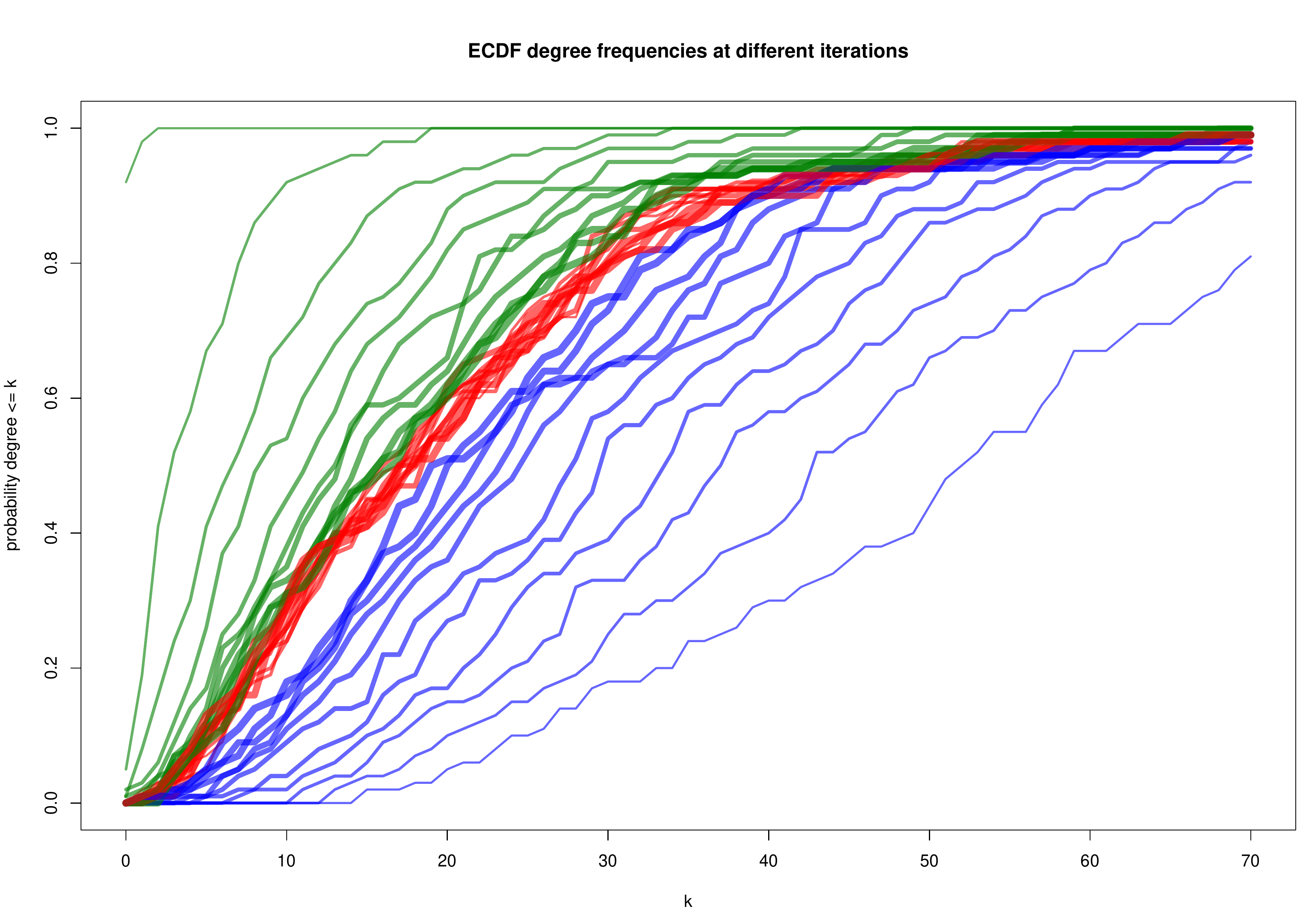}
    \caption{Empirical cumulative distribution function of the degree frequencies, at iterations $1$ (lowest line width), $1001$, $2001$, $...$, $10001$ (largest line width), for the three starting points (stationary - red, sparse - green, dense - blue).}
    \label{fig:MHecdf}
\end{figure}

In Figure \ref{fig:MHecdf} are plotted, for all chains, the empirical cumulative distribution functions of the degree distribution of states at equally spaced iterations.
For the stationary process $(X_t, Z_t)$ these curves, that are plotted in red, are $\{\mathrm{ecdf}(Z_t)\}$ for $t\in\{1,1001,2001,...,10001\}$.
For the processes with sparse and dense starting points, the ECDFs are plotted in blue and green respectively.
The width of the ECDFs is larger when $t$ is as such, showing how the degree distributions of the networks ${X_t}^-$ and ${X_t}^+$ approach the degree distribution of ${X_t}$ as $t\to\infty$.

Markov chains of the type discussed in this application are used to estimate the parameters of models for which the function that normalizes the stationary distribution of the process is not analytic.
For example in \emph{Exponential Random Graph Models} \citep{robins2007introduction} computing the normalization constant requires summing over the set of all possible graphs of a given dimension, making the computation infeasible also for small networks.
Frequentist and Bayesian estimation algorithms for ERGMs \citep{snijders2002markov, caimo2011bayesian} are based on the \emph{Markov Chain Monte Carlo Maximum Liklihood Estimation} method (\emph{MCMCMLE}), developed in \cite{geyer1991markov} and \cite{geyer1992constrained}, where gradients of the likelihood are approximated using a Markov chain that does not require the computation of the normalization constant.

A graph is usually represented by its \emph{adjacency matrix}, with Boolean elements denoting whether two vertices are connected. 
Observed large networks are usually \emph{sparse}, meaning that the number of edges grows linearly with respect to the number of vertices.
If the adjacency matrix is stored as a \emph{dense matrix}, the space required is $O(n^2)$, where $n$ is the size of the network, but individual elements can be updated in constant time ($O(1)$ elementary operations).
Whereas if a \emph{sparse matrix} is used, the space required for storing a sparse network is $O(n)$, but flipping a tie might cost $O(n)$ elementary operations, as the sparse matrix might have to be re-constructed.
The hash table used in \texttt{hset} can be helpful in these cases, as the space required to store the set is $O(n)$, but individual elements are updated with $O(1)$ operations.

\section{Conclusion}
The \texttt{hset} implementation of set operations is motivated by the efficacy of \texttt{hash} data structure when used with reference semantic, allowing significant computational advantages in algorithms in which a set or a multiset is used as container, and it is updated few components at the time.
The implementation is currently restricted to sets or multisets with elements that are numbers (or sets of numbers), so that mathematical relations between integers and reals are respected, e.g., that \texttt{1L} and \texttt{1.0} both represents the number $1$.
This approach differs from the library \texttt{sets} where the classes of two objects determinate whether they can be the same, and reference semantic is not used.

Basic parametrized relations and operations between sets and multisets are implemented, for most of them (all but the intersection) reference semantic can speed up algorithms significantly.
In R, reference semantic is used for \texttt{environment} and \texttt{hash} data objects, whereas almost all other objects use value semantic.
Objects with reference semantic are usually modified by functions with side effects.
In \cite{wickham2019advanced} it is suggested to avoid the use of a function for both its side effects and its returned value.
We partially follow this suggestion, as for the operations defined in Section \ref{subsec:operations} computed with reference semantic, the result of the operation is both returned, and the first operand is transformed to it.
Therefore all terms $\texttt{X1\%|\%X2}$, $\texttt{X1 <- X1\%|\%X2}$, and $\texttt{X1 <- X1\%|}\sim\texttt{\%X2}$ evaluate to the same objects in memory.
The first term is the most common approach to compute operations with reference semantic, where an operator is viewed as accumulator causing the first operand to be modified to the result of the operation with the second operand.
However we suggest using the second approach, that is more coherent with the syntax used with value semantics, and so it is more similar to how code is usually written in R.
Note that the suggestion of avoiding functions with both side effects and returned value is only partially followed, because there are no side effects that modify objects that are not returned.

Our implementation can be useful for simulating Markov chains with countable state space, as in simulation and estimation of temporal network models.
Recent approaches to statistics and probability theory such as \cite{jacobs2019structured}, emphasise the importance of the multiset mathematical structure of discrete probability distributions, especially when determining the properties of algorithms used to sample, estimate or learn probability distributions.
These approaches have been heavily influenced by theories of computation, so they will probably be influential on how algorithms in computational statistics and other disciplines will be developed and implemented.

\section*{Acknowledgement}
Giacomo Ceoldo and Ernst Wit acknowledge funding from the \emph{Swiss National Science Foundation} (SNSF 188534).

\bibliography{biblio.bib}

\appendix 

\section{Semantics in R language}
\label{sec: semantics in programming}
In R, objects are generally modified with \emph{value semantic}, meaning that whenever the object is accessed, a new copy of the object is first created, this new copy is then modified, and eventually stored with the name of the previous variable.
More precisely, a new copy is not always created, because modifying the object locally has computational advantages, this behaviour is called \emph{copy-on-modified} and it is explained in Chapter 2 of \cite{wickham2019advanced}, however when reasoning about the code, it can be assumed that the code behaves as if an object is copied every time it is accessed.
The other approach is to use a \emph{reference semantic}, where the name of the object refers to a pointer to a memory address, so that the modification is always local.
However, operations in the two semantics behave differently. 
For example, in the code
\begin{equation}
\texttt{a = 2;} \ \ 
\texttt{b = a;} \ \
\texttt{b = b + 1;}
\end{equation}
\texttt{a} is initialized to \texttt{2}, then \texttt{b} is defined to be equal to \texttt{a}, and \texttt{b} is incremented.
With value semantic, after the three operations, \texttt{a} and \texttt{b} are equal to \texttt{2} and \texttt{3} respectively, whereas with reference semantic, they are both equal to \texttt{3}.
The reason is that, with value semantic, in the second operation, the value that is referred by \texttt{a} is copied and stored elsewhere, this copy is referred by \texttt{b}, when \texttt{b} is modified, only the value to which \texttt{b} points to is modified, so \texttt{a} has not changed.
On the other hand, if the code above would have been with reference semantic, the second operation would have copied the pointer labelled by \texttt{a}, and this new copied pointer is labelled by \texttt{b}, then \texttt{a} and \texttt{b} would have pointed to the same address in memory, so if \texttt{b} is modified, also \texttt{a} is.

Objects of class \texttt{"numeric"} (as for most types and classes in R), are accessed and updated with value semantic.
Then after the operations above are evaluated, \texttt{a} and \texttt{b} are equal to \texttt{2} and \texttt{3} respectively.
Whereas objects of class \texttt{"hash"} \citep{hash}, are accessed and updated with reference semantic, therefore in the code
\begin{equation}
\begin{split}
&\texttt{a = hash::hash(key="k1", values=2)};\\
&\texttt{b = a;}\\ 
&\texttt{b[["k1"]] = b[["k1"]] + 1};
\end{split}
\end{equation}
\texttt{a} and \texttt{b} refers to the same hash table, so after the last command, \texttt{a[["k1"]]} is also equal to 3.
An hash table can be copied with the following method
\begin{equation}
\begin{split}
&\texttt{a = hash::hash(key="k1", values=2)};\\
&\texttt{b = hash::hash(hash::keys(a), hash::values(a))};\\
&\texttt{b[["k1"]] = b[["k1"]] + 1};
\end{split}
\end{equation}
so that in the second line a new hash table, labelled by \texttt{b} is created, this hash table is a copy of \texttt{a}, as the keys and values are copied from it, but the modification of \texttt{b} in the last row modified, does not change \texttt{a}.

\end{document}